\documentclass[fleqn,usenatbib]{mnras}

\usepackage[T1]{fontenc}

\DeclareRobustCommand{\VAN}[3]{#2}
\let\VANthebibliography\thebibliography
\def\thebibliography{\DeclareRobustCommand{\VAN}[3]{##3}\VANthebibliography}

\usepackage{graphicx}	
\usepackage{amsmath}	
\usepackage{amssymb}	

\usepackage{xcolor}
\usepackage{hyperref}
\newcommand{\orc}{\includegraphics[height=\fontcharht\font`A]{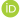}}
\newcommand{\orcid}[1]{\href{https://orcid.org/#1}{\orc}}

\usepackage[normalem]{ulem}

\usepackage{indentfirst}

\usepackage{newtxtext,newtxmath}

\graphicspath{{./}{figures/}}

\def\gsim{\;\rlap{\lower 2.5pt
 \hbox{$\sim$}}\raise 1.5pt\hbox{$>$}\;}
\def\lsim{\;\rlap{\lower 2.5pt
   \hbox{$\sim$}}\raise 1.5pt\hbox{$<$}\;}
\newcommand{\Da}{Damk{\"o}hler }

\title[X-ray to H$\alpha$ SB Ratio Explained]{Clumps in a Cocoon: Geometry and Mixing Set the Universal X-ray to H$\alpha$ Surface Brightness Ratio}

\author[Z. Chen et al.]{
Zirui Chen\orcid{0000-0001-8755-3836}$^{1}$\thanks{E-mail: ziruichen@ucsb.edu}, 
S. Peng Oh\orcid{0000-0002-1013-4657}$^{1}$, 
Drummond B. Fielding\orcid{0000-0003-3806-8548}$^{2}$, 
Lachlan Lancaster\orcid{0000-0002-0041-4356}$^{3,4}$, 
Yuan Li\orcid{0000-0001-5262-6150}$^{5}$,
\newauthor
and Brent Tan\orcid{0000-0003-4805-6807}$^{2}$
\\
$^{1}$Department of Physics, University of California, Santa Barbara, Santa Barbara, CA 93106, USA\\
$^{2}$Department of Physics, New York University, 726 Broadway, New York, NY 10003, USA\\
$^{3}$Center for Computational Astrophysics, Flatiron Institute, 162 5th Ave., New York, NY 10010, USA\\
$^{4}$Department of Astronomy, Columbia University, 550 W 120th St., New York, NY 10025, USA\\
$^{5}$Department of Astronomy, University of Massachusetts, Amherst, MA 01003, USA\\
}

\pubyear{2026}

\begin{document}
\label{firstpage}
\pagerange{\pageref{firstpage}--\pageref{lastpage}}
\maketitle

\begin{abstract}
\noindent Recent observations reveal a nearly universal X-ray to H\(\alpha\) surface-brightness ratio, \({\rm SB_X/SB_{H\alpha}}\sim3\), in galactic winds, ram-pressure-stripped tails, and cluster filaments. This is surprising, given the very different temperatures traced by the X-ray and H$\alpha$ emission. Plane-parallel mixing-layer models do not recover this ratio. We run 3D wind-tunnel simulations in the high density-contrast (\(\chi\sim10^3\)) regime, where the cold phase shatters into small H\(\alpha\)-emitting clumps embedded in a volume-filling X-ray cocoon. After smoothing on the tail-width scale, the simulated ratio converges to the observed value. The hot-gas entrainment rate sets the bolometric luminosity, $L_{\rm bol} \propto \dot{M}_{\rm in}$. H\(\alpha\) carries a nearly fixed fraction of this luminosity, set mainly by atomic emissivity near \(10^4\,{\rm K}\). The X-ray-emitting
mass is the same entrainment rate multiplied by the residence time in the X-ray-emitting temperature band, and its luminosity is the corresponding thermal energy divided by the cooling time. Hence \(L_{\rm X}\propto\dot M_{\rm in}\ \propto L_{\rm bol}\), with the X-ray fraction set by the residence-to-cooling-time ratio. An age diagnostic constrains the residence time to be far shorter than the cooling time, making the X-ray and H\(\alpha\) fractions comparable. Their common scaling with \(L_{\rm bol}\) explains the correlation, while the comparable bolometric fractions explain why the observed ratio is order unity.  
\end{abstract}

\begin{keywords}
turbulence -- galaxies: evolution -- galaxies: structure -- X-rays: galaxies
\end{keywords}



\section{Introduction}\label{sec:intro}

Cold ($\sim 10^4\,{\rm K}$) and hot ($\sim 10^{6}-10^{7}\,{\rm K}$) gas coexist in galactic winds, ram-pressure stripped tails, and cluster filaments. Remarkably, these otherwise very different systems exhibit a similar X-ray to H$\alpha$ surface-brightness ratio, ${\rm SB_X/SB_{H\alpha}}\sim 3$ \citep{Strickland:2002,sun22,Olivares2025}. Because H$\alpha$ traces the cold phase while X-rays trace the hot phase, this correlation provides a rare observational handle on how the two phases are coupled. Any successful explanation---whether based on shocks, conduction, or turbulent mixing---must explain not only the coexistence of the phases, but also the normalization of the ratio.

A natural starting point is turbulent radiative mixing layers. Plane-parallel mixing-layer models \citep{tan21-lines,chen23}, developed and calibrated against numerical simulations \citep{kwak10,ji19,fielding20,tan21}, describe the temperature PDF of a single hot-cold interface. But in the $T_{\rm hot}\sim 10^7\,{\rm K}$ regime relevant here, the predicted X-ray to H$\alpha$ ratio is not robust: using \cite{tan21}, we found that ${\rm SB_X/SB_{H\alpha}}$ varies by $\sim6$ orders of magnitude from $2 \times 10^{-3}$ to $3 \times 10^3$ as we change the temperature dependence of thermal conduction through $\alpha$ ($\kappa \propto T^{\alpha}$) from $\alpha=-1$ (numerical diffusion) to $\alpha=2.5$ (isotropic Spitzer conduction), with ${\rm SB_X/SB_{H\alpha}} \sim 0.02$ for $\alpha=-0.5$ (turbulent diffusion). This misses the observed order-unity value by orders of magnitude unless one fine-tunes $\alpha$. Single-interface models therefore do not robustly explain the observed ratio. This suggests that the central issue is not the detailed microphysics of one interface, but the geometry and temperature PDF of the full multiphase cloud tail.


The key point is that the plane-parallel mixing layer picture is not universally valid \citep{chenoh26geometry}. In wind-tunnel simulations with $T_{\rm hot}\sim 10^6\,{\rm K}$, the cold phase retains a coherent cometary tail, and the temperature PDF is reasonably well described by standard mixing-layer theory. But in the $T_{\rm hot}\sim 10^7\,{\rm K}$ regime relevant for the observed X-ray emission, the density contrast reaches $\chi\equiv \rho_{\rm cold}/\rho_{\rm hot}\sim 10^3$, large enough for the cold phase to shatter into a mist of small clumps \citep{mccourt18,gronke20-mist}. In this regime, the cold phase no longer forms a coherent cloud bounded by a single isolated mixing layer. Instead, the interfaces associated with many small cold clumps overlap and percolate through a connected multiphase volume, so that mixing occurs throughout the cloud tail. The resulting temperature PDF is a collective property of the clump ensemble, rather than a straightforward rescaling of the plane-parallel PDF around a single clump \citep{chenoh26geometry}.

Motivated by this, we run three-dimensional wind-tunnel simulations in the $T_{\rm hot}\sim 10^7\,{\rm K}$, $\chi\sim 10^3$ regime. We show that once the emission is smoothed on scales comparable to the tail width, ${\rm SB_X/SB_{H\alpha}}$ converges to the observed value. The reason is geometric: H$\alpha$ emission arises from many cold clumps, while X-rays arise from a volume-filling cocoon surrounding them. This same picture also explains why the ratio is not enormous. If gas remained in the X-ray band for a full radiative cooling time, it would overproduce X-rays. Instead, turbulent mixing moves gas rapidly through the X-ray-emitting band and into a lower-temperature cooling drain, so the residence time in the X-ray band is much shorter than the radiative cooling time at the characteristic X-ray-emitting temperature.

\section{Numerical Method}\label{sec:numerical_method}

We run three-dimensional wind-tunnel simulations with Athena++ \citep{Stone:2020}, which solves the hydrodynamic equations on a uniform Cartesian grid. We initialize a stationary spherical cloud of radius $R_{\rm cl}=1\,{\rm kpc}$ at $T=10^4\,{\rm K}$ embedded in a hot wind with $T_{\rm hot}=10^7\,{\rm K}$ and Mach number $\mathcal{M}_{\rm wind}=1.5$. The cloud and wind are initially in pressure equilibrium with $P/k_B = 3\times10^4\,{\rm K\,cm^{-3}}$, representative of ICM conditions and consistent with the jellyfish-galaxy simulations of \citet{Lee:2022}. The computational domain has size $250 \,{\rm kpc}\times 15\,{\rm kpc}\times 15\,{\rm kpc}$, with the longest dimension aligned with the wind direction, and is resolved by $4096\times 256\times 256$ cells. This corresponds to a uniform resolution of $60\,{\rm pc}$, so the initial cloud radius is resolved by $\approx 16$ cells. We use outflow boundary conditions on all faces except for a steady inflow imposed in the wind direction. As in \citet{McCourt2015} and \citet{gronke18}, we adopt a cloud-tracking scheme that continuously shifts to the cloud center-of-momentum frame, reducing computational cost and preventing cold gas from leaving the domain. Radiative cooling is implemented with the exact integration scheme of \citet{Townsend2009}, using the collisional-ionization-equilibrium cooling curve of \citet{gnat07} for $T\ge 10^4\,{\rm K}$ and \citet{Koyama:2002} for $T\le 10^4\,{\rm K}$.

We conduct a resolution study by rerunning this fiducial simulation with 2$\times$ higher or lower spatial resolution and show in the middle panel of \autoref{fig:SB_ratio_vs_kernel_size_resolution_time} that our main result, ${\rm SB_X/SB_{H\alpha}}$, is converged against resolution. Additionally, we explore the pressure dependence of ${\rm SB_X/SB_{H\alpha}}$ by rerunning the fiducial simulation with $P/k_B = 10^3$ and $10^6\,{\rm K\,cm^{-3}}$ (representative of galactic wind and cluster-center filaments, respectively) and explore the system size dependence of ${\rm SB_X/SB_{H\alpha}}$ by rerunning the fiducial simulation with 3$\times$ larger initial cloud radius. In these additional runs, we hold the cold- and hot-phase temperatures, the wind Mach number, and the density contrast fixed. Pressure is varied by rescaling the densities of both the initial cloud and the hot wind while maintaining pressure equilibrium, and system size is varied by rescaling the entire simulation domain by the same factor as the initial cloud radius. We discuss the results of these parameter studies in \autoref{sec:physical_origin} and \autoref{sec:robustness}.

To construct surface-brightness maps, we integrate H$\alpha$ and X-ray emissivities along the transverse dimension of the cloud. The H$\alpha$ emissivity is taken from \citet{Ploeckinger20}, and the X-ray emissivity from the Astrophysical Plasma Emission Code \citep[APEC;][]{Smith01}. We adopt a metallicity of $Z=0.5\,Z_\odot$, broadly consistent with \citet{Lee:2022}; varying this to the mean metallicity inferred for the jellyfish tails in \citet{sun22}, $Z\approx 0.18\,Z_\odot$, changes ${\rm SB_X/SB_{H\alpha}}$ only by a factor of order unity. Following \citet{sun22} and \citet{Lee:2022}, we compute X-ray surface brightness in the 0.4--7.5\,keV band, subtract the background wind contribution using the X-ray surface brightness at the edge of the simulation box and then multiply by a bolometric correction factor of $\sim 2$\footnote{The numerical value of the bolometric correction factor is computed from APEC \citep{Smith01} as $c_{\rm bol}=\left. L_{\rm X,bol}\right/L_{\rm X,[0.4, 7.5]keV}$, evaluated at the X-ray emissivity-weighted temperature. We follow \cite{sun22} and \cite{Lee:2022} and use $c_{\rm bol}=2$.} to obtain bolometric X-ray surface brightness. 

\section{Results}
\label{sec:results}

\subsection{Morphology and Smoothing set the X-ray to H$\alpha$ SB Ratio}
\label{sec:sb_smoothing}

\begin{figure}
\centering
\includegraphics[width=\columnwidth]{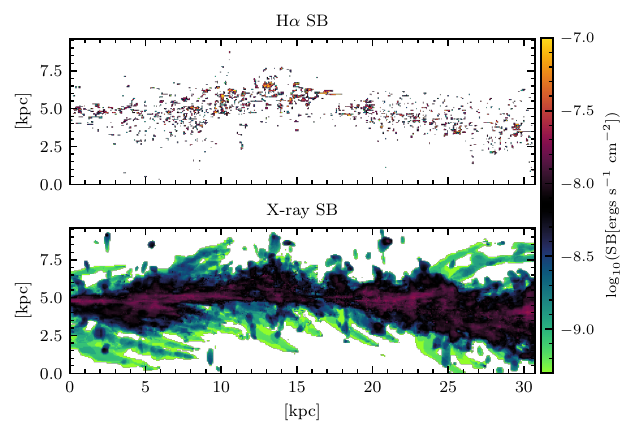}
\caption{H$\alpha$ (\textit{top}) and X-ray (\textit{bottom}) surface brightness maps for a zoomed-in section of the cloud tail in our wind tunnel simulation, where the wind blows from left to right. The H$\alpha$-emitting gas consists of numerous small clumps, while the X-ray-emitting gas is volume-filling and forms an elongated, cocoon-like structure that envelopes the H$\alpha$-emitting clumps. We refer to this morphology as "clumps in a cocoon" throughout this letter.} 
\label{fig:SB_maps_to_show_morphology}
\end{figure}

In \autoref{fig:SB_maps_to_show_morphology}, we show H$\alpha$ (top) and X-ray (bottom) surface brightness maps for a zoomed-in section of our cloud tail. We choose a simulation snapshot where the cloud has undergone its initial evolutionary stages and developed a cometary tail, but the relative shear is still strong, with $\mathcal{M}_{\rm shear} \sim 0.5$. By this stage, the initially $R_{\rm cl}=1\,{\rm kpc}$ cloud has developed a cocoon-like tail of length $\sim 150\,{\rm kpc}$ and transverse diameter $\sim 7\,{\rm kpc}$, i.e. about three times its initial diameter. The cold phase ($T<2\times10^4\,{\rm K}$) has increased in mass by only a factor of $\sim 10$, so most of the cocoon volume is filled not by cold gas but by X-ray-emitting gas at $T\sim {\rm few}\times 10^6\,{\rm K}$. \autoref{fig:SB_maps_to_show_morphology} shows the basic morphological difference between the two tracers: H$\alpha$ emission comes from many small cold clumps, whereas X-ray emission is much smoother and approximately volume-filling across the tail, forming a cocoon that envelopes the H$\alpha$-emitting clumps. This difference matters when measuring ${\rm SB_X/SB_{H\alpha}}$ in finite apertures.

This clumps-in-a-cocoon morphology results from the high density contrast in the $T_{\rm hot}\sim10^7\,{\rm K}$ regime. With $\chi\equiv \rho_{\rm cold}/\rho_{\rm hot}\sim 10^3$, the cold phase shatters into many small clumps rather than remain a monolithic tail\citep{mccourt18,gronke20-mist}\footnote{\cite{gronke20-mist} found that the critical overdensity for cold phase shattering is $\chi\equiv \rho_{\rm cold}/\rho_{\rm hot}\gtrsim 300$.}. By contrast, in the lower-density-contrast $T_{\rm hot}\sim10^6\,{\rm K}$ regime\citep{gronke18}, the cold phase remains coherent and is bounded by a mixing layer, so the temperature PDF is well described by plane-parallel mixing-layer theory (see Figure 3 of \citealt{chenoh26geometry}). In this low density contrast regime, intermediate temperature gas is rare and confined to the thin mixing layer. At high density contrast, however, hot-side isothermal surfaces expand and percolate into connected sheets \citep{chenoh26geometry}; in our simulation these sheets lie near X-ray-emitting temperatures, producing the volume-filling cocoon seen in \autoref{fig:SB_maps_to_show_morphology}. Compared to the planar structure in low-density contrast flows, this greatly increases the amount of X-ray-emitting gas. 

\begin{figure}
\centering
\includegraphics[width=\columnwidth]{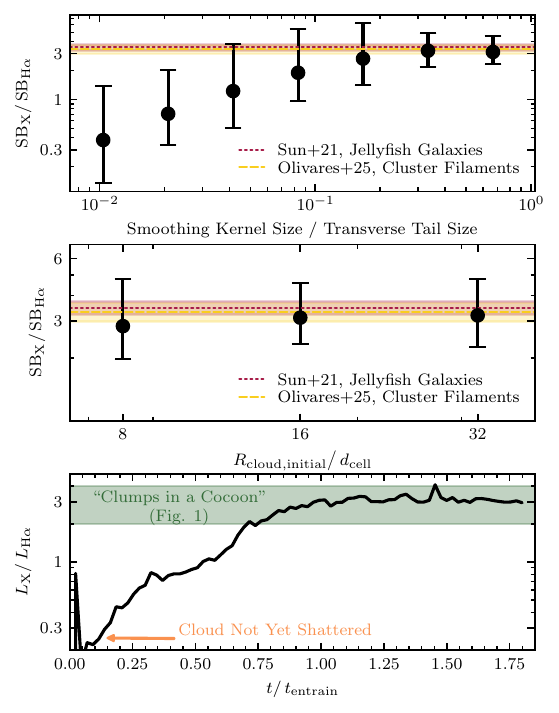}
\caption{\textit{Top}: We smooth the SB maps with Gaussian kernels of varying size and measure ${\rm SB_X/SB_{H\alpha}}$ using apertures with detectable X-ray and H$\alpha$ emission. Error bars show 16th and 84th percentiles. The horizontal bands indicate the observed ratios for jellyfish tails and cluster filaments. As the smoothing scale approaches the transverse tail size, ${\rm SB_X/SB_{H\alpha}}$ rises and saturates at $\sim 3.2$, consistent with observations. \textit{Middle}: Same as the \textit{top} panel, but showing ${\rm SB_X/SB_{H\alpha}}$ versus resolution. We compare our fiducial simulation (16 cells per initial cloud radius) with simulations at $2 \times$ higher and lower resolution, measuring the median ${\rm SB_X/SB_{H\alpha}}$ after smoothing over kernels comparable to the tail width. The ratio is well converged with resolution. \textit{Bottom}: $\left. L_{\rm X}\right/ L_{\rm H\alpha}$ in the cloud tail as a function of time (normalized by the cloud entrainment time $t_{\rm entrain}$). Since surface brightnesses are measured over the same projected tail area, $\left. L_{\rm X}\right/ L_{\rm H\alpha} \sim \left. {\rm SB}_{\rm X}\right/ {\rm SB}_{\rm H\alpha}$. Before cloud shattering, the ratio is small. Once the "clumps in a cocoon" morphology analyzed in \autoref{fig:SB_maps_to_show_morphology} develops (green band), $\left. L_{\rm X}\right/ L_{\rm H\alpha}$ approaches the observed value of $\sim 3$ and remains time-steady after cloud entrainment. We further discuss robustness against cloud evolutionary stage and resolution in \autoref{sec:robustness}. }
\label{fig:SB_ratio_vs_kernel_size_resolution_time}
\end{figure}

To mimic how the data is analyzed in observations, we smooth the surface-brightness maps with Gaussian kernels of varying size and measure ${\rm SB_X/SB_{H\alpha}}$ using only apertures with detectable X-ray and H$\alpha$ emission. This cut is not restrictive for X-rays, but at small smoothing scales it preferentially selects apertures centered on bright H$\alpha$ clumps, biasing ${\rm SB}_{\rm H\alpha}$ high and ${\rm SB_X/SB_{H\alpha}}$ low. As the smoothing scale increases and becomes comparable to the transverse size of the tail, this bias is averaged away, allowing ${\rm SB_X/SB_{H\alpha}}$ to rise and saturate
at $\sim 3.2$, in good agreement with the observed value (top panel of \autoref{fig:SB_ratio_vs_kernel_size_resolution_time}). The key ratio is not the grid-scale ratio, but the coarse-grained ratio measured over apertures large enough to average over the clumpy H$\alpha$ phase.

To test numerical convergence, we run simulations at $2 \times$ higher and lower resolution. The middle panel of \autoref{fig:SB_ratio_vs_kernel_size_resolution_time} shows that ${\rm SB_X/SB_{H\alpha}}$, after smoothing over Gaussian kernels comparable to the tail width, is converged against resolution and agrees well with observations. We further discuss robustness against resolution in \autoref{sec:robustness}.

Notably, ${\rm SB_X/SB_{H\alpha}}$ is not constant throughout cloud evolution, and ${\rm SB_X/SB_{H\alpha}} \sim 3$ emerges dynamically in time. The bottom panel of \autoref{fig:SB_ratio_vs_kernel_size_resolution_time} shows $\left. L_{\rm X}\right/ L_{\rm H\alpha}$, a proxy for $\left. {\rm SB}_{\rm X}\right/ {\rm SB}_{\rm H\alpha}$ (see equation \ref{eq:sb_to_lum}), during cloud evolution. After early fluctuations when the cloud morphology is rapidly evolving, the ratio approaches the observed value of $\sim 3$ once the "clumps in a cocoon" morphology (\autoref{fig:SB_maps_to_show_morphology}) is established, and remains nearly constant after cloud entrainment.

\begin{figure}
\centering
\includegraphics[width=\columnwidth]{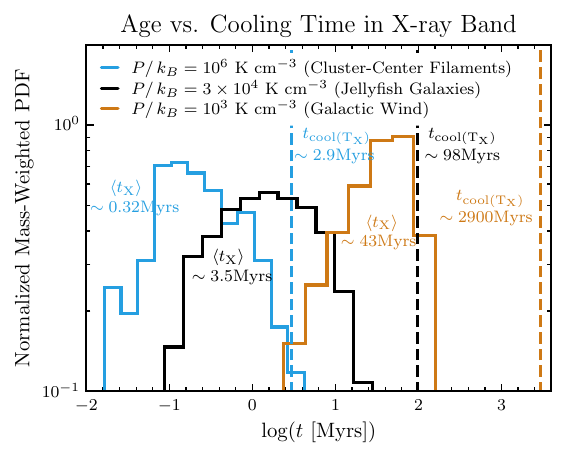}
\caption{Mass-weighted PDFs of the age $t_X$ in the X-ray-emitting temperature band ($10^6\,{\rm K}<T<8\times10^6\,{\rm K}$), which contains $\simeq 90\%$ of the X-ray emission. The age is measured using an passive scalar that advects with the flow and is incremented by $\Delta t$ every timestep while gas remains in the X-ray band and reset to zero upon exit. Brown, black, and blue show runs with $P/k_B=10^3,\ 3\times10^4$, and $10^6\ {\rm K\,cm^{-3}}$, representative of galactic winds, jellyfish tails, and cluster filaments, respectively.Vertical dashed lines mark $t_{\rm cool}(T_X)$. In all cases, $\langle t_X \rangle \ll t_{\rm cool}(T_X)$, indicating that turbulent mixing processes gas through the X-ray-emitting band before it cools radiatively. The systematic shift of the PDFs with pressure suggests that 
$t_X$ decreases with pressure, leaving ${\rm SB_X/SB_{H\alpha}}$ only weakly pressure-dependent.}
\label{fig:residence_time_vs_cooling_time_in_X_ray_emitting_band}
\end{figure}

\subsection{Physical origin of the X-ray to H$\alpha$ SB ratio}
\label{sec:physical_origin}

Thus far, we have shown that our simulations reproduce the observed ${\rm SB_X/SB_{H\alpha}}$. However, these results do not by themselves explain why the ratio is of order unity. To gain physical insight, we now develop a simple order-of-magnitude explanation for the numerical value of ${\rm SB_X/SB_{H\alpha}}$. In a plane-parallel mixing layer, the natural estimate is given by $
{{\rm SB_X}}/{{\rm SB_{H\alpha}}}
\sim
[{n_{\rm X}^2\Lambda_{\rm X} \ell_{\rm X}}]/[{n_{\rm H\alpha}^2\Lambda_{\rm H\alpha}\ell_{\rm H\alpha}}]$. This framing is appropriate for a single interface, where both emitting phases reside in thin sheets. In the clumps-in-a-cocoon morphology established in \autoref{fig:SB_maps_to_show_morphology}, however, there is no unique pair of emitting layers: H$\alpha$ arises from many cold clumps, while X-rays arise from a volume-filling cocoon. The more useful description is therefore in terms of total luminosities, or equivalently the temperature PDF and relative emitting volumes.

For the cloud tail as a whole, the measured surface brightnesses can be regarded as luminosities averaged over the same projected smoothing aperture $A_{\rm ap}$, so that
\begin{align}
    \frac{{\rm SB}_{\rm X}}{{\rm SB}_{\rm H\alpha}}
    \sim
        \frac{L_{\rm X}/A_{\rm ap}}{L_{\rm H\alpha}/A_{\rm ap}}
    \sim
    \frac{L_{\rm X}}{L_{\rm H\alpha}}.
    \label{eq:sb_to_lum}
\end{align}
Note that \autoref{eq:sb_to_lum} is only true on large spatial scales once $A_{\rm ap}$ is sufficiently large to average out the distinct geometry of X-ray and H$\alpha$ emitting gas.
The importance of the clumps-in-a-cocoon morphology is not this relation itself, but that it strongly boosts $L_{\rm X}$. In a plane-parallel mixing layer, X-ray-emitting gas is confined to a thin hot-side sheet and is correspondingly rare. In our shattered-tail regime, by contrast, X-ray-emitting gas becomes a volume-filling cocoon surrounding the cold clumps, greatly increasing the abundance of gas near X-ray-emitting temperatures and hence the X-ray luminosity.

We now relate both luminosities to the bolometric luminosity $L_{\rm bol}$. On the H$\alpha$ side,
\begin{align}
    \frac{L_{\rm H\alpha}}{L_{\rm bol}}
    \sim
    \frac{\int n^2 \Lambda_{\rm H\alpha} dV}{\int n^2 \Lambda_{\rm bol} dV}
    \sim
    \left.\frac{\Lambda_{\rm H\alpha}}{\Lambda_{\rm bol}}\right|_{T\sim 10^4\,{\rm K}}
    \equiv
    f_{\rm H\alpha}
    \sim
    5\times 10^{-3}.
    \label{eq:Lha_over_Lbol}
\end{align}
This follows because both \(dL_{\rm H\alpha}/d\log T\) and \(dL_{\rm bol}/d\log T\) are sharply peaked near \(T\sim10^4\,{\rm K}\), reflecting the corresponding peaks in \(n^2\Lambda_{\rm H\alpha}\) and \(n^2\Lambda_{\rm bol}\) under approximately isobaric conditions. Both luminosity integrals are therefore dominated by gas near this temperature. We have verified that \autoref{eq:Lha_over_Lbol} holds well in all our simulations.

In \autoref{eq:Lha_over_Lbol}, we expressed $L_{\rm bol}$ as the volume integral of the bolometric cooling emissivity. Alternatively, we can estimate $L_{\rm bol} \sim h_{\rm hot}\dot{M}_{\rm in}$ since in steady state the energy source powering the radiation is the inflowing hot gas enthalpy \citep{begelman90,ji19}. Here, $h_{\rm hot}\simeq (5/2)k_B T_{\rm hot}/\mu m_p$ is the hot-phase enthalpy per unit mass, where $T_{\rm hot}=10^7$K as in our simulations, and $\dot{M}_{\rm in}$ is the mass inflow rate onto the cloud. Rewriting $L_{\rm bol}$ allows us to relate it to $L_{\rm X}$ as
\begin{align}
    \frac{L_{\rm X}}{L_{\rm bol}}
    \sim
    \frac{n_{\rm X}^2 \Lambda_{\rm X}(T_{\rm X}) V_{\rm X}}
         {h_{\rm hot}\dot{M}_{\rm in}}
    \sim
    \frac{(h_{\rm X} \rho_{\rm X}/t_{\rm cool}(T_{\rm X})) V_{\rm X}}
         {h_{\rm hot}\dot{M}_{\rm in}}
    \sim
    \frac{h_{\rm X} M_{\rm X}/t_{\rm cool}(T_{\rm X})}
         {h_{\rm hot}\dot{M}_{\rm in}},
    \label{eq:Lx_over_Lbol_basic}
\end{align}
where $M_{\rm X}$ is the mass of X-ray-emitting gas, and $h_{\rm X}\simeq (5/2)k_B T_{\rm X}/\mu m_p$ is the enthalpy per unit mass for X-ray-emitting gas, analogous to $h_{\rm hot}$. $T_{\rm X}$ is the characteristic X-ray temperature calculated using the emission weighted geometric mean 
\begin{align}
    \log_{10} T_{\rm X}
    =
    \frac{\sum (\epsilon_{\rm X}-\epsilon_{\rm X,wind}) \log_{10} T}
         {\sum (\epsilon_{\rm X}-\epsilon_{\rm X,wind})},
    \label{eq:Tx_def}
\end{align}
where the X-ray emissivities are measured in the 0.4-7.5 keV band, and we subtract the wind contribution in the same way as for the surface-brightness maps. This gives $T_{\rm X}\sim 3\times 10^6\,{\rm K}$, and $t_{\rm cool}(T_{\rm X})$ provides an estimate of the characteristic cooling time for X-ray-emitting gas. Writing
\begin{align}
    M_{\rm X}\sim \dot{M}_{\rm in} \langle t_{\rm X,res} \rangle,
\end{align}
where $\langle t_{\rm X,res} \rangle$ is the mean residence time in the X-ray-emitting band, gives
\begin{align}
    \frac{L_{\rm X}}{L_{\rm bol}}
    \sim
    \frac{h_{\rm X}}{h_{\rm hot}}
    \frac{\langle t_{\rm X,res} \rangle}{t_{\rm cool}(T_{\rm X})}
    \sim
    \frac{T_{\rm X}}{T_{\rm hot}}
    \frac{\langle t_{\rm X,res} \rangle}{t_{\rm cool}(T_{\rm X})}.
    \label{eq:Lx_over_Lbol_tx}
\end{align}

We now estimate the mean residence time $\langle t_{\rm X,res} \rangle$. We define the X-ray-emitting band to be $10^6\,{\rm K}<T<8\times10^6\,{\rm K}$, which contains $\simeq 90\%$ of the total X-ray emission. We use a passive scalar that advects with the flow and is initialized to zero throughout the simulation domain. At each timestep, this scalar is incremented by $\Delta t$ in cells whose cell-averaged temperature lies in the X-ray-emitting band, and is reset to zero otherwise. Thus, for gas currently in the X-ray-emitting band, the scalar measures an age $t_{\rm X}$: the continuous time since its most recent entry into the band. The age $t_X$ is a biased estimator of the Lagrangian residence time $t_{\rm X,res}$, since long-lived events are over-represented in any given snapshot. However, in steady state, one can show robustly that the mean residence time $\langle t_{\rm X,res} \rangle$ and mean age $\langle t_X \rangle$ obey\footnote{Consider an event of length $t$. Its observed age $t_X$ is uniformly distributed between 0 and $t$, so that $\langle t_X | t_{\rm X,res}=t \rangle =t/2$. At any given snapshot, the observed age distribution is time-weighted (longer events are more likely to be observed), $p_{\rm snapshot}(t) = t p(t)/\langle t_{\rm X,res} \rangle$, where $p(t)$ is the true residence time distribution. Thus, $\langle t_{\rm X} \rangle = \int (t/2) p_{\rm snapshot}(t) dt = \langle t_{\rm X,res}^2 \rangle/2 \langle t_{\rm X,res} \rangle = [\langle t_{\rm X,res} \rangle/2] (1 + \sigma_{\rm t_{\rm X,res}}^2/\langle t_{\rm X,res} \rangle^2)$, where $\sigma_{\rm t_{\rm X,res}}$ is the standard deviation of the $t_{\rm X,res}$ distribution, and this relationship implies $\langle t_{\rm X,res} \rangle \le 2 \langle t_{\rm X} \rangle$.}: 
\begin{equation}
\langle t_{\rm X,res} \rangle \le 2 \langle t_X \rangle 
\end{equation}
so we can bound the true mean residence time $\langle t_{\rm X,res} \rangle$ from above. Since both the passive scalar and gas energy are mixed conservatively, numerical diffusion does not introduce a differential bias between the cell temperature and the mean age $\langle t_X \rangle$. In what follows, we use this mean age $\langle t_X \rangle$ to approximate the mean residence time $\langle t_{\rm X,res} \rangle$ appearing in \autoref{eq:Lx_over_Lbol_tx}.

\autoref{fig:residence_time_vs_cooling_time_in_X_ray_emitting_band} shows the resulting mass-weighted PDFs of $t_{\rm X}$ for runs with $P/k_B=10^3$, $3\times10^4$, and $10^6\,{\rm K\,cm^{-3}}$, representative of galactic winds, jellyfish galaxies, and cluster-center filaments, respectively. The corresponding mass-weighted means are $\langle t_{\rm X}\rangle \simeq 43$, $3.5$, and $0.32\,{\rm Myr}$. These values are all much shorter than the cooling time at the characteristic X-ray temperature, which are $t_{\rm cool}(T_{\rm X})\simeq 2900$, $98$, and $2.9\,{\rm Myr}$ in the three runs. Thus the gas currently emitting in X-rays has typically not remained continuously in the X-ray-emitting temperature band for a full cooling time at $T_{\rm X}$.

Combining \autoref{eq:Lha_over_Lbol} and \autoref{eq:Lx_over_Lbol_tx}, using values from our fiducial simulation with $P/k_B=3\times10^4\,{\rm K\,cm^{-3}}$ (appropriate for jellyfish galaxy conditions), and assuming $\langle t_{\rm X,res}\rangle\sim\langle t_{\rm X}\rangle$, we obtain
\begin{align}
    \frac{L_{\rm X}}{L_{\rm H\alpha}}
    \sim
    \frac{1}{f_{\rm H\alpha}}
    \frac{T_{\rm X}}{T_{\rm hot}}
    \frac{\langle t_{\rm X} \rangle}{t_{\rm cool}(T_{\rm X})}
    \sim
    \frac{1}{5\times 10^{-3}}
    \frac{3\times 10^6\,{\rm K}}{10^7\,{\rm K}}
    \frac{3.5\,{\rm Myr}}{98\,{\rm Myr}}
    \sim
    2,
    \label{eq:Lx_over_Lha}
\end{align}
in good agreement with the saturated ${\rm SB}_{\rm X}/{\rm SB}_{\rm H\alpha}$ measured in the simulation and observations.

The key physical result is that the X-ray-emitting gas is not a long-lived reservoir cooling radiatively at $T_{\rm X}$. If gas remained in the X-ray band for a full cooling time at $T_{\rm X}$, it would overproduce X-rays. Instead, turbulent mixing processes gas rapidly through the X-ray-emitting band and into a lower-temperature cooling drain. The abundance of X-ray-emitting gas is therefore regulated by the temperature cascade and its associated mass flux, not simply by radiative cooling at $T_{\rm X}$ itself \footnote{Gas evolves isobarically in the simulations. Adiabatic processes at constant pressure have fixed density and temperature. Temperature can only evolve via entropy-changing processes such as radiative cooling and mixing.}.



The pressure dependence in \autoref{fig:residence_time_vs_cooling_time_in_X_ray_emitting_band} is particularly suggestive. Lowering the pressure by a factor of \(30\) from the fiducial value increases \(\langle t_{\rm X}\rangle\) by a factor of \(\sim 12\), while increasing the pressure by a comparable factor decreases it by a factor of \(\sim 11\). Across these three simulations, this corresponds approximately to \(\langle t_{\rm X}\rangle\propto P^{-0.7}\), intermediate between a pressure-independent dynamical time and the \(P^{-1}\) scaling of the cooling time at fixed temperature. Moreover, \(\langle t_{\rm X}\rangle\) is of the same order as the cooling time near \(\widetilde{T}\sim\sqrt{T_hT_c}\sim10^{5.5}\,{\rm K}\), a characteristic mixed-gas temperature in a strongly cooling part of the cooling curve \citep{begelman90}. Together, these results suggest that passage through the X-ray band is jointly regulated by mixing and radiative cooling, perhaps through mixing into a lower-temperature cooling outlet, although the present simulations do not uniquely identify the controlling process. We therefore regard the X-ray band as a transient stage in a mixing-driven cooling flow rather than a long-lived reservoir.

Since $t_{\rm cool}(T_{\rm X})$ falls with pressure, the ratio $\langle t_{\rm X} \rangle/t_{\rm cool}(T_{\rm X})$ varies only weakly with pressure (roughly $\propto P^{0.3}$ in our limited sample of 3); so does $L_{\rm X}/L_{\rm bol}$ in \autoref{eq:Lx_over_Lbol_tx}. In our runs with $P/k_B=10^3$, $3\times10^4$, and $10^6\,{\rm K\,cm^{-3}}$, we have $L_{\rm X}/L_{\rm H\alpha}= 1.2, 3.0$, and 8.2, respectively. This helps explain why ${\rm SB}_{\rm X}/{\rm SB}_{\rm H\alpha}$ remains of order unity across different environments, even if it is not strictly pressure-independent. This is also qualitatively consistent with the observations of \citet{sun22}, who found that the SB ratio in jellyfish tails varies only modestly, from $\sim 2.5$ to $\sim 4.4$, across clusters with different pressure profiles. Their sample spans a much narrower pressure range than our simulations, so at present this should be viewed only as suggestive evidence for a weak pressure dependence rather than a precise scaling. \cite{chenoh26geometry} showed that the temperature PDF of a multiphase flow is controlled by a set of dimensionless parameters. Changing pressure changes $t_{\rm cool}$ and thus the dimensionless \Da number (Da=$t_{\rm mix} / t_{\rm cool}$, \citealt{damkohler40, tan21}), so naturally the temperature PDF changes and affects ${\rm SB}_{\rm X}/{\rm SB}_{\rm H\alpha}$. A broader and more quantitative exploration of pressure in both simulations and observations is left to future work.

This picture is supported by a second passive scalar that tracks gas that has ever cooled below $2\times10^4\,{\rm K}$. Within the X-ray-emitting band, the mass-weighted mean concentration is $\sim 0.5$, implying that roughly half the mass in this band previously cooled below $2 \times 10^4$K. The abundance of X-ray-emitting gas is therefore regulated by mixing with the cold phase, not by its own radiative cooling time.

\subsection{Robustness to size, evolutionary stage, and resolution}
\label{sec:robustness}

The pressure dependence of ${\rm SB}_{\rm X}/{\rm SB}_{\rm H\alpha}$ is discussed in \autoref{sec:physical_origin}, where we showed that the key quantity is the ratio $t_{\rm X}/t_{\rm cool}(T_{\rm X})$, which is only weakly pressure dependent. Here we briefly summarize the remaining robustness checks.

\textit{System size:}
The universal X-ray to H$\alpha$ SB ratio is observed across a wide range of scales. In our picture, this is expected because both $L_{\rm X}$ and $L_{\rm H\alpha}$ scale approximately with the cocoon volume: the X-ray-emitting gas is volume-filling, while the H$\alpha$ luminosity can be written schematically as $L_{\rm H\alpha}=N_{\rm clump}L_{\rm clump}$, with the number of cold clumps proportional to the cocoon volume. In an additional run with an initial cloud radius three times larger, the coarse-grained surface-brightness ratio again approaches \({\rm SB}_{\rm X}/{\rm SB}_{\rm H\alpha}\sim3\), despite a factor $\sim 27$ change in volume.

\textit{Evolutionary stage and shear:}
As shown in the bottom panel of \autoref{fig:SB_ratio_vs_kernel_size_resolution_time}, ${\rm SB}_{\rm X}/{\rm SB}_{\rm H\alpha}$ evolves during the early stages but asymptotes to $\sim 3$ once the clumps-in-a-cocoon morphology is established, remaining nearly constant through cloud entrainment. This is consistent with galactic wind observations where the ratio persists out to several kpc from the disk \citep{Strickland:2002}. The saturation associated with dynamical entrainment is reflected in the turbulent velocity field. During the early shearing phase, the X-ray-emitting gas has a larger turbulent velocity than the H$\alpha$-emitting gas, reflecting stronger coupling to the hot wind. After entrainment, the two converge, indicating that both tracers sample the same residual turbulent motions.

\textit{Resolution:}
Recent JWST observations resolve $\sim 5-15 {\rm pc}$ PAH-emitting clouds in M82 \citep{Bolatto2024, Fisher2025}. While qualitatively consistent with our fragmented cold phase, these scales are unresolved in our simulations, which have a fiducial resolution of $\sim 60 {\rm pc}$. Many cold clumps are close to the grid scale in our simulations, and we do not resolve the cooling length, $c_s t_{\rm cool}$ evaluated at the cool gas temperature \citep{mccourt18}. However, the integrated H$\alpha$ luminosity depends primarily on the total amount of gas near $T\sim 10^4\,{\rm K}$, not on the size of individual clumps:
\begin{align}
    L_{\rm H\alpha}
    =
    \int \frac{dV}{d\log T}\,\epsilon_{\rm H\alpha}(T)\,d\log T. \label{eq:Lha_from_PDF}
\end{align}
We verified this using simulations at $2\times$ higher and lower spatial resolution, which leave the volume-weighted temperature PDF and $L_{\rm X}, L_{\rm H \alpha}$ (not just their ratio) unchanged. Consequently, ${\rm SB}_{\rm X}/{\rm SB}_{\rm H\alpha}$ is converged (middle panel of \autoref{fig:SB_ratio_vs_kernel_size_resolution_time}), consistent with previous work showing that cold gas mass converges at far lower resolution than cold gas morphology \citep[see][for a review]{FGOH23}. We therefore do not claim convergence of the clump-size spectrum, only of the emission-weighted quantities relevant for this Letter.
 
\section{Discussion}

\subsection{Comparison with previous simulations and observations}

One of the first quantitative measurements of the X-ray to H$\alpha$ surface-brightness ratio in a multiphase outflow was made for the halo of NGC\,253, where \citet{Strickland:2002} found ${\rm SB_X/SB_{H\alpha}}\sim 0.5$--$2.2$. More recently, \citet{sun22} found a tight relation in jellyfish tails, ${\rm SB_X/SB_{H\alpha}}=3.48\pm0.25$, while \citet{Olivares2025} reported a similar correlation in cluster filaments. Our simulation reproduces this observed order-unity ratio once the emission is smoothed over scales comparable to the tail width, consistent with the fact that the observational apertures are much larger than individual cold clumps. 

On the theoretical side, \citet{Lee:2022} found ${\rm SB_X/SB_{H\alpha}}\sim 1$--$20$ in galaxy-scale stripping simulations, with substantial scatter at native resolution that is reduced after coarse-graining over large regions. This is qualitatively consistent with our interpretation that the measured ratio depends on aperture scale when H$\alpha$ emission is clumpy. \citet{Tonnesen11} obtained similar stripped-tail X-ray/H$\alpha$ ratios; their three runs imply ratios of $\sim 1.3$--$4.6$. These works support the view that the observed ratio is not unique to one numerical setup, but emerges generically in multiphase stripped tails.

What this paper adds is a physical explanation for why the ratio is order unity. In the plane-parallel mixing-layer picture, the X-ray-emitting gas occupies only a thin hot-side sheet, and the predicted ratio is far too small. In our shattered-tail regime, by contrast, the X-ray-emitting gas becomes volume-filling and forms a cocoon around the cold clumps, strongly boosting the abundance of gas near X-ray-emitting temperatures. At the same time, \autoref{fig:residence_time_vs_cooling_time_in_X_ray_emitting_band} shows that this gas does not remain in the X-ray band for a full cooling time: turbulent mixing processes it onward into a lower-temperature cooling drain. The observed ratio therefore reflects both geometry and mixing-regulated residence times.

\subsection{Caveats and broader implications}

\textit{Conduction.}
We do not include explicit thermal conduction in our fiducial simulations, so we cannot yet determine how the shattered-tail solution is modified by conductive transport. However, auxiliary plane-parallel mixing-layer tests using the analytic model in \cite{tan21} show that the X-ray to H$\alpha$ ratio is extremely sensitive to the assumed temperature dependence of the conductivity, which is controlled by the power-law exponent $\alpha$ in $\kappa \propto T^{\alpha}$. Without explicit conduction ($\alpha=-1$ for numerical diffusion), such layers underproduce X-rays, yielding ${\rm SB_X/SB_{H\alpha}}\sim 2 \times 10^{-3}$, whereas isotropic Spitzer conduction ($\alpha=2.5$) overproduces them, yielding ${\rm SB_X/SB_{H\alpha}}\sim 3\times10^3$. Between these cases, ${\rm SB_X/SB_{H\alpha}}$ monotonically increases with $\alpha$, and $\alpha=0.80$ reproduces the observed ratio of $\sim 3$, but this requires fine-tuning and does not by itself explain the observed robustness across environments. These conclusions apply only to plane-parallel models. Given the importance of geometry in our results, a Spitzer-like conductivity could still be consistent with the observed ${\rm SB_X/SB_{H\alpha}}$ ratio in complex geometries. Conduction is likely important in some systems. \citet{Rodriguez2026} concluded that conduction explains the centrally concentrated X-ray emission in \textit{Chandra} observations of 30 Doradus relative to the wind-blown bubble simulations of \citet{Lancaster25b}. We therefore regard conduction as an important uncertainty in the temperature PDF, but not a compelling alternative to our geometric and mixing-based explanation. Testing anisotropic conduction in the shattered-tail regime is an important target for future work.

\textit{Magnetic fields.}
Magnetic fields could alter the cold phase morphology. However, recent turbulent-box studies suggest that this need not strongly affect the global mass exchange. Hydrodynamic simulations of multiphase turbulence produce exponential cold-gas growth and a fog of small droplets \citep{gronke22-turb}, while MHD simulations find only marginal differences in cold-gas growth and survival despite significantly more filamentary morphology \citep{das24}. This is encouraging for our picture: the X-ray/H$\alpha$ ratio may depend more on the overall abundance of cold and X-ray-emitting gas than on the exact topology of individual cold structures.


\textit{Charge exchange.} The APEC emissivities used here assume collisional ionization equilibrium and omit charge-exchange emission, which may contribute to soft X-rays at hot–cold interfaces \citep{lallement04,lallement12}, and could be important in starburst galaxies \citep{Liu12, Zhang14, Cumbee16, Fukushima24}. Its magnitude in multiphase flows is uncertain and sensitive to unresolved interface structure and non-equilibrium ion populations. Although our simulations reproduce the observed X-ray/H\(\alpha\) ratio without this process, we cannot rule out a significant charge exchange contribution in observed systems. Such emission could modify the ratio, and quantifying its importance requires further investigation.

More broadly, the lesson of this work is that X-ray luminosity in turbulent multiphase gas need not trace the radiative cooling time at the X-ray-emitting temperature, but can instead be limited by how rapidly mixing transfers hot gas into a lower-temperature cooling outlet. A similar idea has been invoked by \citet{metzger25}, who argued that turbulent mixing with cool gas can strongly suppress shock-powered thermal X-rays in novae by reducing the thickness and volume of the hot emitting layer. The universal X-ray to H$\alpha$ ratio therefore suggests that mixing, geometry, and residence times jointly regulate multiphase emission.

\section{Conclusions}

We have argued that the nearly universal X-ray to H$\alpha$ surface-brightness ratio, ${\rm SB}_{\rm X}/{\rm SB}_{\rm H\alpha}\sim 3$, is best understood not in terms of a single plane-parallel mixing layer, but through the geometry and mass processing of a shattered multiphase cloud tail. In the high density contrast regime, the cold phase fragments into many small H$\alpha$-emitting clumps surrounded by a volume-filling X-ray-emitting cocoon. This morphology boosts the abundance of X-ray-emitting gas relative to plane-parallel expectations, while turbulent mixing prevents that gas from lingering in the X-ray band for a full cooling time and thereby overproducing X-rays. The measured pressure trend suggests joint regulation by mixing and radiative cooling, although the controlling process is not uniquely identified.  More broadly, this exercise shows that X-ray luminosity in multiphase turbulence need not trace either local shock heating or the cooling time at the X-ray temperature itself; it can instead be controlled by how rapidly mixing transfers gas through the hot phase and into a lower-temperature drain. The observed X-ray/H$\alpha$ ratio thus provides a promising empirical probe of temperature PDFs, emitting volumes, and mass-processing rates in multiphase gas. An important next step will be to test how this picture is modified by anisotropic conduction, magnetic fields, and charge exchange emission in the shattered-tail regime.

\section*{Acknowledgements}

We thank the anonymous referee for insightful comments that improved the quality of this work. We thank the organizers of the KITP program "Turbulence in Astrophysical Environments", where this collaboration was formed and some of the ideas were discussed. We thank Yuanyuan Su and Taya Govereen-Segal for helpful conversations. Zirui Chen and Peng Oh acknowledge NSF grant AST240752 and HST grant AR-17860 for support. Drummond B. Fielding acknowledges support from NSF through grant AST-2601424, NASA through grants HST-AR-17859.015-A and HST-AR17559.009-A, and the Simons Foundation through the Simons Initiative on the Geometry of Flows (Grant Award ID BDTargeted-00017375, DF). Yuan Li acknowledges support from NASA grant 80NSSC22K0668, Chandra X-ray Observatory grant TM3-24005X, NSF grant AST-2510198, and CAREER award AST-2516092. This work made considerable use of the Stampede3 supercomputer through allocation TG-PHY240194 from the Advanced Cyberinfrastructure Coordination Ecosystem: Services \& Support (ACCESS) program, which is supported by National Science Foundation grants \#2138259, \#2138286, \#2138307, \#2137603, and \#2138296.

\section*{Data Availability}

The data underlying this article will be shared upon reasonable request to the corresponding author.



\bibliographystyle{mnras}
\bibliography{reference, master_references}




\appendix


\bsp	
\label{lastpage}
\end{document}